\documentclass[11pt]{article}
\usepackage{epsfig}
\usepackage{cite}
\def\d{\hbox{d}}

\def\ln{\hbox{ln}}

\begin{document}
\unitlength1cm
\begin{titlepage}
\vspace*{-1cm}
\begin{flushright}
ZU-TH 04/08
\end{flushright}
\vskip 3.5cm

\begin{center}
{\Large\bf Matching NLLA+NNLO for event shape distributions}
\vskip 1.cm
{\large  T.~Gehrmann}$^a$, {\large G.~Luisoni}$^a$ and
{\large H.\ Stenzel}$^b$
\vskip .7cm
{\it $^a$
Institut f\"ur Theoretische Physik, Universit\"at Z\"urich,
Winterthurerstrasse 190,\\ CH-8057 Z\"urich, Switzerland}
\vskip .3cm
{\it $^b$
II.\ Physikalisches Institut, Justus-Liebig Universit\"at Giessen\\
Heinrich-Buff Ring 16, D-35392 Giessen, Germany
}
\end{center}
\vskip 2cm

\begin{abstract}
We study the matching of the next-to-leading logarithmic
approximation (NLLA) onto the  fixed
next-to-next-to-leading order (NNLO) calculation for event
shape distributions in electron-positron annihilation. The
resulting theoretical predictions combine all precision QCD
knowledge on the distributions, and are theoretically
reliable over an extended kinematical range. Compared to
previously available matched NLLA+NLO and fixed order NNLO
results, we observe that the effects of the combined
NLLA+NNLO are small in the three-jet region, relevant for
precision physics.
\end{abstract}
\vskip 3cm
\texttt{Keywords:} QCD, Jets, higher order
calculations, resummation. \vfill
\end{titlepage}
\newpage

Event shape distributions
in $e^+e^-$ annihilation processes
are classical hadronic observables which can be measured very accurately
and provide an ideal proving ground for testing our understanding of strong
interactions.
The deviation from simple two-jet configurations, which are
a limiting case in event shapes,  is proportional
to the strong coupling constant, so that
by comparing the measured
event shape distribution  with the theoretical
predictions, one can determine the strong coupling constant
$\alpha_{s}$. At LEP, a standard set of event shapes was studied in
great detail~\cite{aleph,opal,l3,delphi}:
thrust $T$~\cite{farhi} (which is substituted here
by $\tau = 1-T$),
heavy jet mass $\rho$~\cite{mh},
wide and total jet broadening $B_W$ and $B_T$~\cite{bwbt},
$C$-parameter~\cite{c} and two-to-three-jet transition parameter
in the Durham algorithm $Y_3$~\cite{durham}.
The definitions of
these variables, which we denote collectively
as $y$ in the following,  are summarised in~\cite{hasko}.
The two-jet limit of
each  variable is $y\to 0$.

The theoretical prediction is made within perturbative QCD, expanded to a
finite order in the coupling constant. This fixed order expansion
is reliable only if the event shape variable is sufficiently far
away from its two-jet limit. In the approach to this limit, event shapes
display large infrared logarithms at all orders in perturbation theory, such
that the expansion in the strong coupling constant fails to converge.
Resummation of these logarithms yields a
description appropriate to the two-jet limit. To explain event shape
distributions over their full kinematical range, both descriptions need to be
matched onto each other.
Until very recently, the theoretical state-of-the-art description of
event shape distributions was based on the matching of the
next-to-leading-logarithmic approximation
(NLLA,~\cite{resumall}) onto the
fixed next-to-leading order (NLO,~\cite{ERT,kunszt,event}) calculation.
Using the newly available results of the  next-to-next-to-leading order (NNLO)
corrections for the standard set of event
shapes~\cite{ourt,our3j,ourevent,ourjets}
 introduced above, we
derive here matching of the resummed NLLA onto the fixed order NNLO.

For two-particle final states, all above event shape
variables  have the fixed value $y=0$, consequently their
distributions receive their first non-trivial contribution
from three-particle final states, which, at order
$\alpha_s$, correspond to three-parton final states.
Therefore, both theoretically and experimentally, these
distributions are closely related to three-jet production.

Fixed-order QCD corrections to event shape distributions were
calculated long ago to next-to-leading order
(NLO,~\cite{ERT,kunszt,event}), and most recently
to next-to-next-to-leading order (NNLO,~\cite{ourt,our3j,ourevent}).
At a centre-of-mass energy $Q$ and for renormalisation scale $\mu$,
they take the form:
\begin{eqnarray}
\frac{1}{\sigma_{{\rm had}}}\, \frac{\d\sigma}{\d y}
(y,Q,\mu) &=& \bar\alpha_s (\mu) \frac{\d \bar{A}}{\d y}(y)
+ \bar\alpha_s^2 (\mu) \frac{\d \bar{B}}{\d y} (y,x_\mu) +
\bar\alpha_s^3 (\mu) \frac{\d \bar{C}}{\d y}(y,x_\mu) +
{\cal O}(\bar\alpha_s^4)\;, \label{eq:NNLOmu}
\end{eqnarray}
where
\begin{equation}
  \bar\alpha_s = \frac{\alpha_s}{2\pi}\;, \qquad
x_\mu = \frac{\mu}{Q}\;,
\end{equation}
and
where $\bar A$, $\bar B$ and $\bar C$ are the perturbatively calculated
coefficients at LO, NLO, NNLO, normalised to $\sigma_{{\rm had}}$,
explicit relations are given in~\cite{ourevent} (note the different convention for the $\beta_{i}$ coefficients).

The resummation of large logarithmic corrections in the $y\to 0$ limit
starts from the integrated cross section:
\begin{equation}\label{Rfixed}
R\left(y,Q,\mu\right)\,\equiv\,\frac{1}{\sigma_{{\rm had}}}\int_{0}^{y}\frac{d\sigma\left(x,Q,\mu\right)}{dx}dx,
\end{equation}
which has the following fixed-order expansion:
\begin{equation}\label{Rfixedexp}
R\left(y,Q,\mu\right)\,=\,1+\,\mathcal{A}\left(y\right)
\bar{\alpha}_{s}\left(\mu\right)\,
+\,\mathcal{B}\left(y,x_\mu\right)\bar{\alpha}_{s}^{2}\left(\mu\right)\,+\,\mathcal{C}\left(y,x_\mu\right)\bar{\alpha}_{s}^{3}\left(\mu\right).
\end{equation}
The fixed-order coefficients $\mathcal{A}$, $\mathcal{B}$, $\mathcal{C}$
can be obtained by integrating the distribution (\ref{eq:NNLOmu})
and using $R(y_{{\rm max}},Q,\mu)=1$ to all orders, where
$y_{{\rm max}}$ is the maximal
kinematically allowed value for the shape variable
$y$.

In the limit $y\to 0$ one observes that the perturbative
$\alpha_{s}^n$--contribution to $R(y)$ diverges like $\alpha_s^n L^{2n}$,
with $L=-\ln\, y$ ($L=-\ln\,(y/6)$ for $y=C$).
This leading logarithmic (LL) behaviour is due to multiple soft gluon
emission at higher orders, and the LL coefficients exponentiate, such that
\begin{displaymath}
\ln\,  R(y) \sim L g_1(\alpha_s L)\;,
\end{displaymath}
where $g_1(\alpha_s L)$ is a power series in its argument.

For the event shapes considered here,
leading and  next-to-leading logarithmic
(NLL) corrections can be resummed to all orders in the coupling constant,
such that
\begin{equation}\label{eq:Rresummed}
R\left(y,Q,\mu\right)\,=\ \left(1+C_{1}\bar{\alpha}_{s}
\right)\,e^{\left(L\,g_{1}\left(\alpha_{s}L\right)
+g_{2}\left(\alpha_{s}L\right)\right)}\;,
\end{equation}
where terms beyond NLL have been consistently omitted, and
$\mu=Q$ ($x_{\mu}=1$) is used. In the case of the
C-parameter further large logarithms around
$C\,\approx\,0.75$, produce a so-called Sudakov shoulder in
the distribution due to soft gluon divergences within the
physical region \cite{resumcshoulder}.

By differentiating expression (\ref{eq:Rresummed}) with
respect to $y$, one recovers the resummed differential
event shape distributions, which yield an accurate
description for $y\to 0$. The first complete calculation
of next-to-next-to-leading logarithmic (NNLL) corrections
to event shape distributions is available  for the
energy-energy correlation function~\cite{eecdg}, which is
not part of the standard set of event shape observables.
The application of soft-collinear effective field theory to
event shape distributions~\cite{scet} promises to yield
results beyond NLL. Most recently, this formalism was
applied to compute the resummed thrust distribution beyond
NLL accuracy~\cite{scetthrust}.

Closed analytic forms for the  LL and NLL resummation
functions $g_1(\alpha_s L)$, $g_2(\alpha_s L)$ are
available for $\tau$~\cite{resumt}, $\rho$~\cite{resumrho},
$B_W$ and $B_T$~\cite{resumbwbt,resumbwbtrecoil},
$C$~\cite{resumc} and $Y_3$~\cite{resumy3a}. For the
convenience of the reader, we collect them in uniform
notation in an Appendix. They can be expanded as power
series, such that:
\begin{equation}
\ln R(y,Q,\mu) = \sum_{i=1}^{\infty}\sum_{n=1}^{i+1}
G_{i,i+2-n} \bar{\alpha}_s^i L^{i+2-n}\;,
\label{eq:gexpand}
\end{equation}

To obtain a reliable description of the event shape
distributions over a wide range in $y$, it is mandatory to
combine fixed order and resummed predictions. To avoid the
double counting of terms common to both, the two
predictions have to be matched onto each other. A number of
different matching procedures have been proposed in the
literature, see for example~\cite{hasko} for a review. The
by-now standard procedure is the so-called $\ln\,
R$-matching~\cite{resumall}. In this particular scheme, all
matching coefficients can be extracted analytically from
the resummed calculation, while most other schemes require
the numerical extraction of some of the matching
coefficients from the distributions at fixed order. Since
the fixed order calculations face numerical instabilities
in the region $y\to 0$, these matching coefficients can
often be determined only within large errors. We shall
therefore consider only the $\ln\, R$-matching here. The
$\ln\, R$-matching at NLO is described in detail
in~\cite{resumall}, where the authors also anticipated the
fixed-order NNLO corrections to be available shortly, and
briefly outlined this matching scheme to NNLO.

In the $\ln\, R$-matching scheme, the NLLA+NNLO expression is
\begin{eqnarray}\label{logRmatching}
\ln\left(R\left(y,\alpha_{S}\right)\right)&=&L\,g_{1}\left(\alpha_{s}L\right)\,+\,g_{2}\left(\alpha_{s}L\right)\nonumber\\
&&+\,\bar{\alpha}_{S}\left(\mathcal{A}\left(y\right)-G_{11}L-G_{12}L^{2}\right)+{}\nonumber\\
&&+\,\bar{\alpha}_{S}^{2}\left(\mathcal{B}\left(y\right)-\frac{1}{2}\mathcal{A}^{2}\left(y\right)-G_{22}L^{2}-G_{23}L^{3}\right){}\nonumber\\
&&+\,\bar{\alpha}_{S}^{3}\left(\mathcal{C}\left(y\right)-\mathcal{A}\left(y\right)\mathcal{B}\left(y\right)+\frac{1}{3}\mathcal{A}^{3}\left(y\right)-G_{33}L^{3}-G_{34}L^{4}\right)\;.
\end{eqnarray}
The matching coefficients appearing in this expression can
be obtained from (\ref{eq:gexpand}) and  are listed in
Table~\ref{tab:coeff}. In the matching of $Y_3$, the
constants ${\cal F}_i$ depend on the jet
algorithm~\cite{resumy3a}, in general, they can be
determined only numerically. For the Durham-algorithm, one
finds  ${\cal F}_2=-\pi^2/32$ and ${\cal F}_3=0.0620\pm
0.0100$~\cite{giulia}, using the semi-numerical resummation
method described in~\cite{resumy3b}. Numerical values of
the matching coefficients for $N=3$, $N_F=5$ are given in
Table~\ref{tab:numericalcoeff}.

To ensure the vanishing of the matched expression
at the kinematical boundary $y_{\textrm{\tiny{max}}}$, the further
substitution~\cite{hasko} is made:
\begin{equation}\label{Ltilde}
L\,\longrightarrow\,\tilde{L}\,
=\,\frac{1}{p}\,\ln\left(\left(\frac{y_{0}}{x_{L}\,y}\right)^{p}
-\left(\frac{y_{0}}{x_{L}\,y_{\textrm{\tiny{max}}}}\right)^{p}+1\right),
\end{equation}
where $y_0 = 6$ for $y=C$ and $y_0=1$ otherwise. $p=1$ and
$x_{L}=1$ is taken as default.

The full renormalisation scale dependence of (\ref{logRmatching}) is
given by replacing the coupling constant, the fixed-order coefficients,
the resummation functions and the matching coefficients as follows:
\begin{eqnarray}
\alpha_s & \to & \alpha_s(\mu)\;, \\
\nonumber \\
\mathcal{B}\left(y\right) &\to &
\mathcal{B}\left(y,\mu\right)=2\,\beta_{0}\,
\ln x_\mu \, \mathcal{A}\left(y\right)
+\mathcal{B}\left(y\right)\;,\nonumber \\
\mathcal{C}\left(y\right) & \to &
\mathcal{C}\left(y,\mu\right)=\left(2\,\beta_{0}\,
\ln x_\mu \right)^{2}\mathcal{A}\left(y\right)
+2\,\ln x_\mu \,\left[2\,\beta_{0}
\mathcal{B}\left(y\right)+2\,\beta_{1}\,\mathcal{A}\left(y\right)\right]
+\mathcal{C}\left(y\right)\;,
\label{fixedorderrenscaledependence}\\
\nonumber\\
g_2\left(\alpha_{S}L\right) &\to &
{g}_{2}\left(\alpha_{S}L,\mu^{2}\right)
=g_{2}\left(\alpha_{S}L\right)+\frac{\beta_{0}}{\pi}
\left(\alpha_{S}L\right)^{2}\,
g_{1}'\left(\alpha_{S}L\right)\,\ln x_\mu \;,
\label{g2mudep} \\
\nonumber \\
G_{22}&\to & G_{22}\left(\mu\right)=G_{22}\,+\,2\beta_{0}G_{12}\ln x_\mu
\;,\nonumber\\
G_{33}&\to & {G}_{33}\left(\mu\right)=G_{33}\,
+\,4 \beta_{0}  G_{23}\ln x_\mu\,.
\label{Gijdeponrenorm}
\end{eqnarray}
In the above, $g_1'$ denotes the derivative of $g_1$ with
respect to its argument. The LO coefficient ${\cal A}$ and
the LL resummation function $g_1$, as well as the matching
coefficients $G_{i\,i+1}$ remain independent on $\mu$.

The arbitrariness in the choice of the logarithm to be
resummed can be quantified by varying the constant $x_{L}$.
This variation implies also the modification of the NLL
resummation function and of its coefficients
\begin{eqnarray}
g_{2}\left(\alpha_{S}L\right)\rightarrow\,\tilde{g}_{2}(\alpha_{S}\tilde{L})&=&g_{2}(\alpha_{S}\tilde{L})+\frac{d}{d\tilde{L}}\left(\tilde{L}g_{1}(\alpha_{S}\tilde{L})\right)\ln x_{L}\,,\\
G_{11}\,\rightarrow\,\tilde{G}_{11}&=&G_{11}+2G_{12}\ln x_{L}\,,\nonumber \\
G_{22}\,\rightarrow\,\tilde{G}_{22}&=&G_{22}+3G_{23}\ln x_{L}\,,\nonumber \\
G_{33}\,\rightarrow\,\tilde{G}_{33}&=&G_{33}+4G_{34}\ln
x_{L}\,.
\end{eqnarray}

\begin{figure}[t]
\centering
  \includegraphics[width=1.0\textwidth]{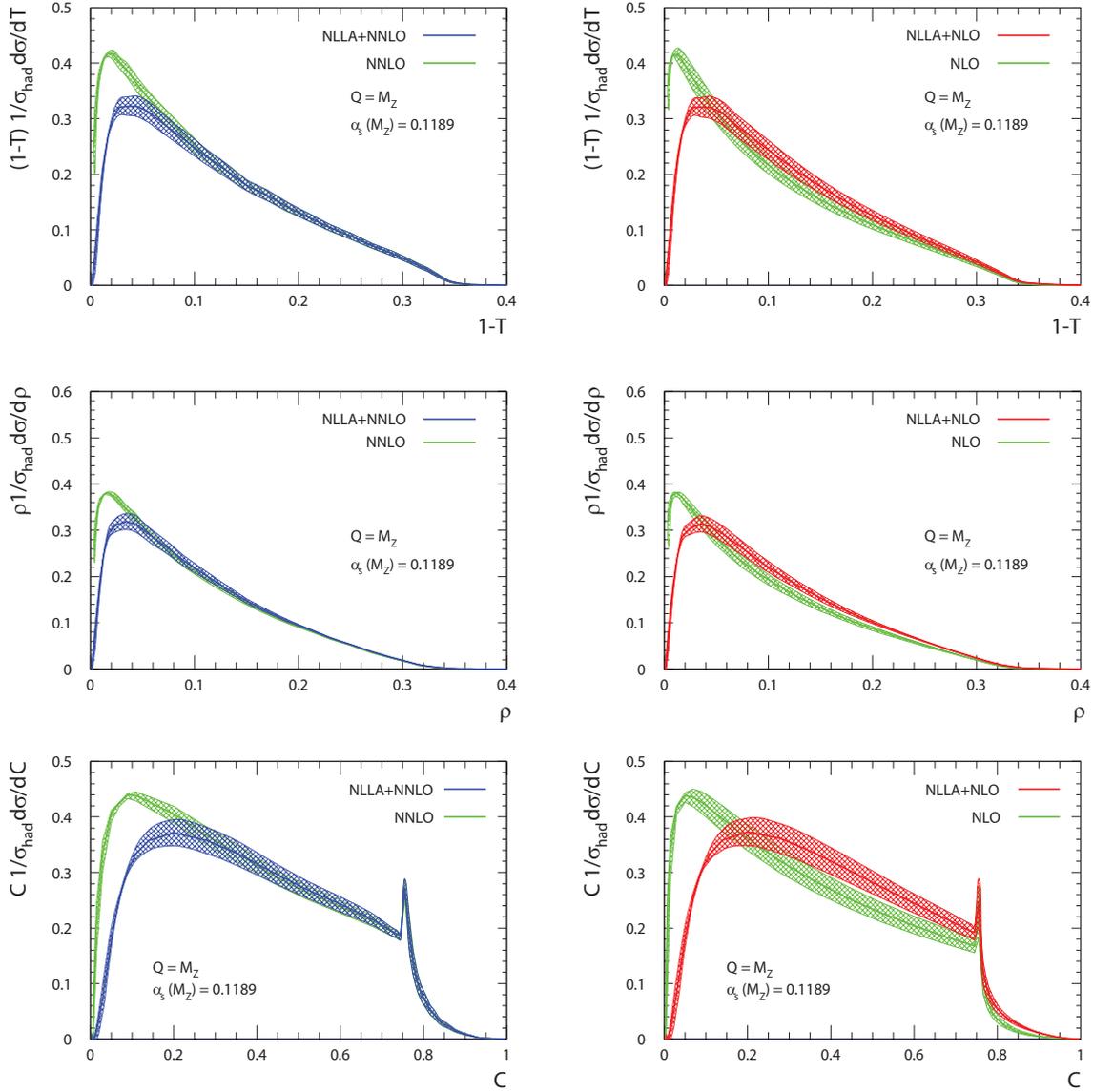}\\
  \caption{Comparison of the matched NLLA+NNLO and NLLA+NLO with fixed order NNLO and NLO predictions for the thrustlike observables $\tau$, $\rho$ and $C$-parameter.}\label{fig:thrustlikeplot}
\end{figure}

\begin{figure}[t]
\centering
  \includegraphics[width=1.0\textwidth]{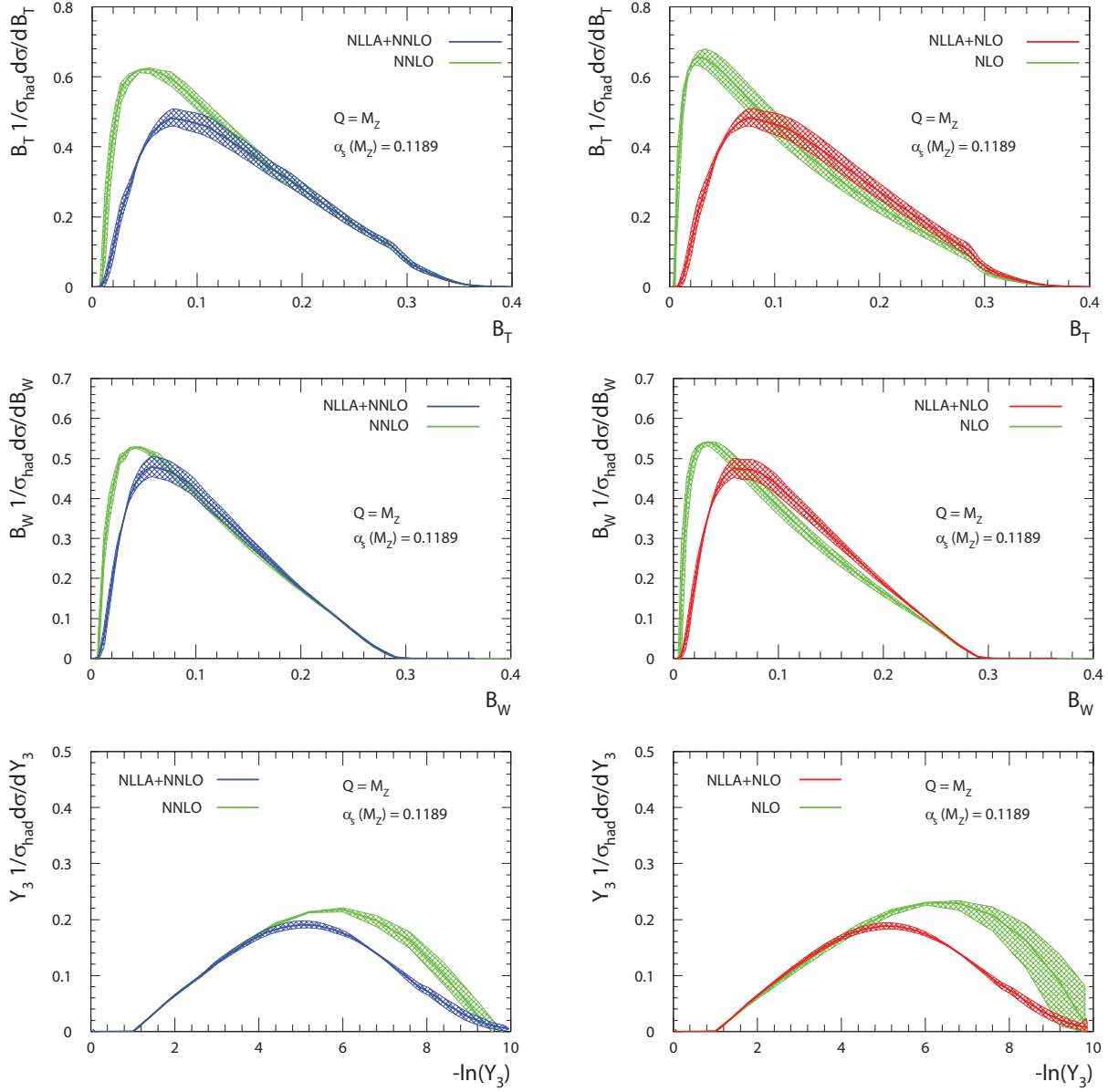}\\
  \caption{Comparison of the matched NLLA+NNLO and NLLA+NLO with fixed order NNLO and NLO predictions for $B_{T}$, $B_{W}$ and $Y_{3}$.}\label{fig:broadeningsplot}
\end{figure}

In Figures \ref{fig:thrustlikeplot} and
\ref{fig:broadeningsplot}, we compare the matched NLLA+NNLO
predictions for all event shape variables with the fixed
order NNLO predictions, and the matched NLLA+NLO with fixed
order NLO. To allow for a better distinction of the
different descriptions, all distributions were weighted by
the respective shape variables. We use $Q=M_Z$ and fix
$x_\mu = 1$, the strong coupling constant is taken as the
current world average $\alpha_s(M_Z) =
0.1189$~\cite{bethke}. To quantify the renormalisation
scale uncertainty, we have varied $1/2<x_\mu<2$, resulting
in the error band on these figures.

Several common effects are seen for all shape variables. The most
striking observation is that the difference between NLLA+NNLO
and NNLO is largely restricted to the two-jet region, while
NLLA+NLO differ in normalisation throughout the full kinematical range.
This behaviour may serve as a first indication for the
numerical smallness of corrections beyond NNLO in the three-jet region.

An immediate consequence of this behaviour concerns the extraction of
$\alpha_s$ from event shape data. Studies at LEP~\cite{aleph,delphi,opal,l3}
yielded substantially different values (by about 10-15\%) from
 NLO and NLLA+NLO theory. This discrepancy is an immediate consequence of the
varying normalisations in the two approaches.
One can expect
that $\alpha_s$ obtained using
NLLA+NNLO will differ from the fixed-order NNLO result~\cite{ouralphas}
only moderately, given the good agreement of both descriptions in the three-jet
region for fixed  $\alpha_s$.

In the approach to the two-jet region, the NLLA+NLO and NLLA+NNLO
predictions agree by construction, since the matching suppresses any
fixed order terms. Equally, the renormalisation scale uncertainty on
both these predictions is identical in this region. In the three-jet region,
 NLLA+NNLO agrees with NNLO. The difference between NLLA+NNLO and
NLLA+NLO is only moderate in the three-jet region, and
especially much smaller than the difference between the
fixed order NNLO and NLO predictions. The renormalisation
scale uncertainty in the three-jet region is reduced by
20-40\% between NLLA+NLO and NLLA+NNLO.

\begin{figure}[t]
\centering
  \includegraphics[width=1.0\textwidth]{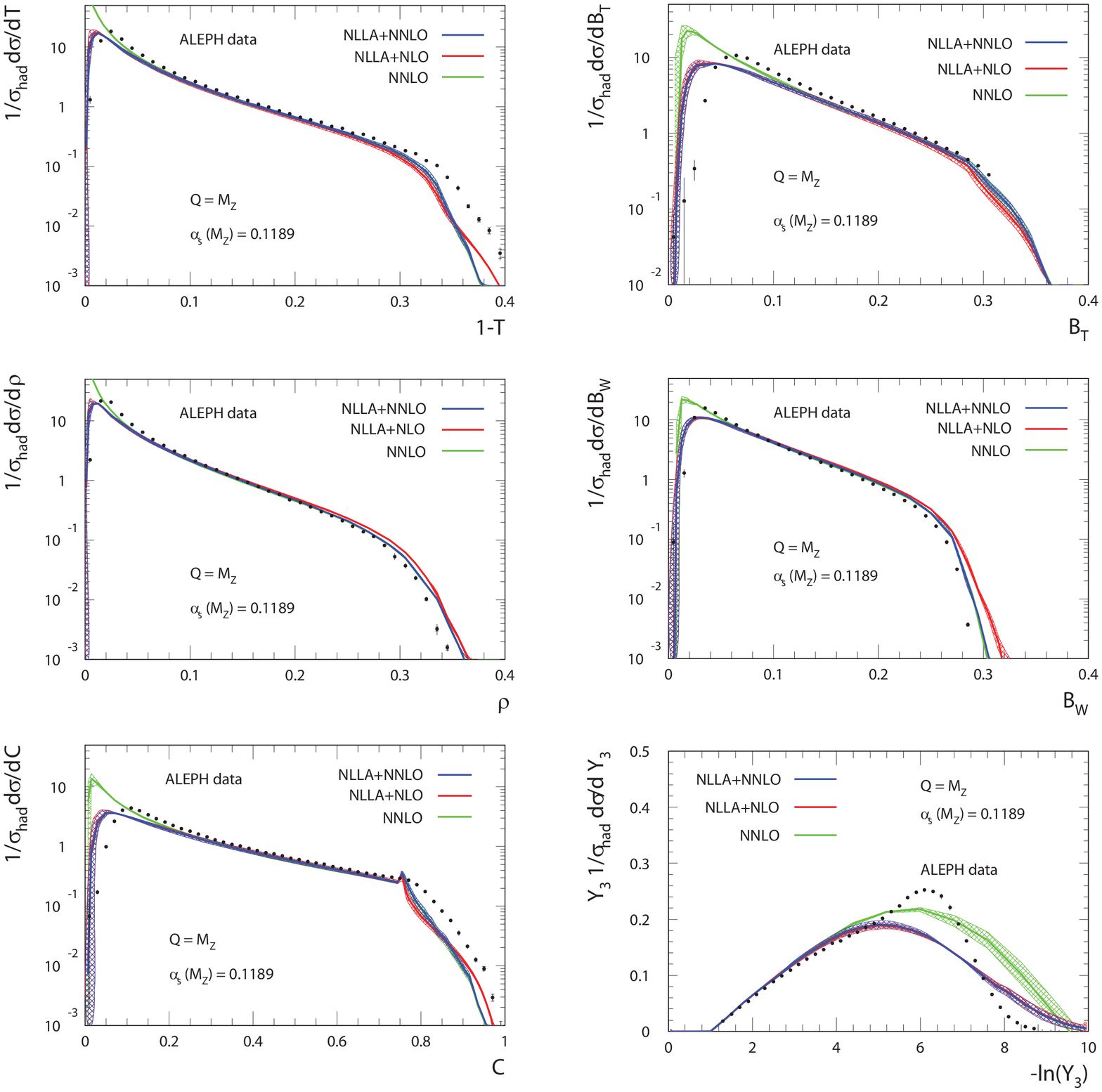}\\
  \caption{Comparison of the matched NLLA+NNLO and NLLA+NLO with fixed order NNLO with the hadron-level data taken by the ALEPH experiment~\cite{aleph}.}\label{fig:data}
\end{figure}

The parton-level fixed order NNLO and matched NLLA+NLO and NLLA+NNLO
predictions are compared
to hadron-level data taken by the ALEPH experiment~\cite{aleph}
in Figure~\ref{fig:data}.
The description of the hadron-level data improves between
parton-level NLLA+NLO and parton-level NLLA+NNLO, especially
in the three-jet region for most event shapes. The behaviour in the
two-jet region is described better by the resummed predictions than by the
fixed order NNLO, although the agreement is far from perfect.
This discrepancy was observed already
in earlier studies based on  NLLA+NLO. It can in part be
attributed to hadronisation corrections, which become
large in the approach to the two-jet limit. A very recent study of
logarithmic corrections beyond NLLA for the thrust
distribution~\cite{scetthrust} also shows that subleading
logarithms in the two-jet region can account for about half of this
discrepancy.

A precise extraction of $\alpha_s$ from event shape data
will require the inclusion of hadronisation corrections
and of quark mass effects (at least to NLO~\cite{quarkmass}), as
done already in the fixed order NNLO study~\cite{ouralphas}.
It can be anticipated that inclusion of the matched NLLA+NNLO
corrections results in a further improvement of the extraction
of $\alpha_s$ from event shape data over results obtained previously
at NLLA+NLO as well as at NNLO. The principal shortcomings of the up-to-now
default  NLLA+NLO studies were the substantial  renormalisation scale
uncertainty and the sizable scatter of values of $\alpha_s$ obtained
from different shape variables.
 It was observed recently, that a fixed-order
NNLO extraction~\cite{ouralphas}
 reduces the renormalisation scale uncertainty
by a factor 1.3 compared to NLLA+NLO and eliminates the
scatter between different observables. It will be very
interesting to see the impact of the matched NLLA+NNLO
calculation on the extraction of $\alpha_s$. We will
address this issue in a future study.

A routine implementing the matching for all event shapes discussed here
can be obtained upon request from the authors.

\section*{Acknowledgements}
We would like to thank Giulia Zanderighi and Thomas
Becher for useful discussions.
This research was supported by the Swiss National Science Foundation
(SNF) under contract 200020-117602.

\appendix
\section{Resummation Functions}
We summarize here the expressions for
the resummed NLL
integrated cross section (\ref{eq:Rresummed}) for different event
shapes.
One has
\begin{displaymath}
R\left(y,Q,\mu\right)\,=\,\left(1+C_{1}\bar{\alpha}_{s}
\right)\,\Sigma\left(y\right)\;,
\end{displaymath}
with
$$\Sigma\left(y\right)\,=\,\exp\left\{Lg_{1}\left(\alpha_{S}L\right)+g_{2}\left(\alpha_{S}L\right)\right\}.$$

Following~\cite{resumbwbtrecoil,resumy3a},
and in order to
unify the notation,
the resummed part is then expressed
through auxiliary functions
$h_{1}\left(\lambda\right)$ and
$h_{2}\left(\lambda\right)$, with:
\begin{displaymath}
\Sigma\left(y\right)\,=\,\Sigma_{s}\left(y\right)\,\mathcal{F}\left(R'\right)
\end{displaymath}
where
$$R'\left(\lambda\right)=-\frac{1}{2}\left[h_{1}\left(\lambda\right)+\lambda\,h_{1}'\left(\lambda\right)\right].$$
The functions $h_{1}\left(\lambda\right)$,
$h_{2}\left(\lambda\right)$, $\Sigma_{s}\left(y\right)$ and
$\mathcal{F}\left(R'\right)$ depend on the event shape
observable, as well as the parameter $\lambda$. The QCD
constants $\beta_{0}$, $\beta_{1}$ and $K$ are normalised
as follows:
\begin{eqnarray*}
\beta_{0}&=&\frac{1}{12}\left(11C_{A}-2N_{F}\right)\;,\\
\beta_{1}&=&\frac{1}{24}\left(17\,C_{A}^{2}-5\,C_{A}N_{F}-3C_{F}N_{F}
\right)\;,\\
K&=&C_{A}\left(\frac{67}{18}-\frac{\pi^{2}}{6}\right)-\frac{5}{9}\,N_{F}\,.
\end{eqnarray*}
\subsection{Thrust and C-Parameter}
From \cite{resumt} and \cite{resumc}, one has:
\begin{eqnarray*}
\lambda&=&\frac{\beta_{0}}{\pi}\,\alpha_{S}L\;,\\
h_{1}\left(\lambda\right)&=&-\frac{C_{F}}{2\lambda\beta_{0}}\left[\left(1-2\,\lambda\right)\ln\left(1-2\lambda\right)-2\left(1-\lambda\right)\ln\left(1-\lambda\right)\right]\;,\\
h_{2}\left(\lambda\right)&=&-\frac{C_{F}\,K}{4\beta_{0}^{2}}\left[2\,\ln\left(1-\lambda\right)-\ln\left(1-2\lambda\right)\right]-\frac{3C_{F}}{4\beta_{0}}\ln\left(1-\lambda\right)\nonumber\\
&&-\frac{C_{F}\beta_{1}}{2\beta_{0}^{3}}\left(\ln\left(1-2\lambda\right)-2\,\ln\left(1-\lambda\right)+\frac{1}{2}\ln^{2}\left(1-2\lambda\right)-\ln^{2}\left(1-\lambda\right)\right)\;,\\
\Sigma_{s}\left(y\right)&=&e^{L\,2h_{1}\left(\lambda\right)+2h_{2}\left(\lambda\right)}\;,\\
\mathcal{F}\left(R'\right)&=&\frac{e^{-2\gamma_{E}R'}}{\Gamma\left(1+4\,R'\right)}\,.
\end{eqnarray*}
These yield:
\begin{eqnarray*}
g_{1}\left(\alpha_{S}L\right)&=&2\,h_{1}\left(\frac{\beta_{0}}{\pi}\,\alpha_{S}L\right)\;,\\
g_{2}\left(\alpha_{S}L\right)&=&2\,h_{2}\left(\frac{\beta_{0}}{\pi}\,\alpha_{S}L\right)-\ln\left[\Gamma\left(1+4R'\right)\right]-2\gamma_{E}R'\,.
\end{eqnarray*}
\subsection{Heavy Jet Mass}
From~\cite{resumt} one has:
\begin{eqnarray*}
\lambda&=&\frac{\beta_{0}}{\pi}\,\alpha_{S}L\;,\\
h_{1}\left(\lambda\right)&=&-\frac{C_{F}}{2\lambda\beta_{0}}\left[\left(1-2\,\lambda\right)\ln\left(1-2\lambda\right)-2\left(1-\lambda\right)\ln\left(1-\lambda\right)\right]\;,\\
h_{2}\left(\lambda\right)&=&-\frac{C_{F}\,K}{4\beta_{0}^{2}}\left[2\,\ln\left(1-\lambda\right)-\ln\left(1-2\lambda\right)\right]-\frac{3C_{F}}{4\beta_{0}}\ln\left(1-\lambda\right)\nonumber\\
&&-\frac{C_{F}\beta_{1}}{2\beta_{0}^{3}}\left(\ln\left(1-2\lambda\right)-2\,\ln\left(1-\lambda\right)+\frac{1}{2}\ln^{2}\left(1-2\lambda\right)-\ln^{2}\left(1-\lambda\right)\right)\;,\\
\Sigma_{s}\left(y\right)&=&e^{L\,2h_{1}\left(\lambda\right)+2h_{2}\left(\lambda\right)}\;,\\
\mathcal{F}\left(R'\right)&=&\frac{e^{-2\gamma_{E}R'}}{\Gamma\left(1+2\,R'\right)^{2}}\;.
\end{eqnarray*}
These yield:
\begin{eqnarray*}
g_{1}\left(\alpha_{S}L\right)&=&2\,h_{1}\left(\frac{\beta_{0}}{\pi}\,\alpha_{S}L\right)\;,\\
g_{2}\left(\alpha_{S}L\right)&=&2\,h_{2}\left(\frac{\beta_{0}}{\pi}\,\alpha_{S}L\right)-2\,\ln\left[\Gamma\left(1+2R'\right)\right]-2\gamma_{E}R'\,.
\end{eqnarray*}
\subsection{Total Jet Broadening}
From \cite{resumbwbt,resumbwbtrecoil} one has:
\begin{eqnarray*}
\lambda&=&2\,\frac{\beta_{0}}{\pi}\,\alpha_{S}L\;,\\
h_{1}\left(\lambda\right)&=&\frac{2\,C_{F}}{\lambda\beta_{0}}\left(\ln\left(1-\lambda\right)+\lambda\right)\;,\\
h_{2}\left(\lambda\right)&=&-\frac{C_{F}\,K}{2\beta_{0}^{2}}\left(\ln\left(1-\lambda\right)+\frac{\lambda}{1-\lambda}\right)-\frac{3\,C_{F}}{2\beta_{0}}\ln\left(1-\lambda\right)\nonumber\\
&&+\frac{C_{F}\beta_{1}}{\beta_{0}^{3}}\left(\frac{1}{2}\ln^{2}\left(1-\lambda\right)+\frac{\ln\left(1-\lambda\right)}{1-\lambda}+\frac{\lambda}{1-\lambda}\right)\;,\\
\Sigma_{s}\left(y\right)&=&e^{L\,h_{1}\left(\lambda\right)+h_{2}\left(\lambda\right)}\;,\\
\mathcal{F}\left(R'\right)&=&\left[\int_{1}^{\infty}\frac{dx}{x^{2}}\left(\frac{1+x}{4}\right)^{-R'}\right]^{2}\frac{e^{-2\gamma_{E}R'}}{\Gamma\left(1+2\,R'\right)}\nonumber\\
&=&\left[\frac{4^{R'}\;{}_{2}F_{1}\left(R',\,1+R';\,2+R';\,-1\right)}{\left(1+R'\right)}\right]^{2}\frac{e^{-2\gamma_{E}R'}}{\Gamma\left(1+2\,R'\right)}\;.
\end{eqnarray*}
These yield:
\begin{eqnarray*}
g_{1}\left(\alpha_{S}L\right)&=&\,h_{1}\left(\frac{\beta_{0}}{\pi}\,\alpha_{S}L\right)\;,\\
g_{2}\left(\alpha_{S}L\right)&=&\,h_{2}\left(\frac{\beta_{0}}{\pi}\,\alpha_{S}L\right)-\ln\left[\Gamma\left(1+2R'\right)\right]-2\gamma_{E}R'\nonumber\\
&&+2\,\ln\left[\frac{4^{R'}\;{}_{2}F_{1}\left(R',\,1+R';\,2+R';\,-1\right)}{\left(1+R'\right)}\right]\;.
\end{eqnarray*}
\subsection{Wide Jet Broadening}
From \cite{resumbwbt,resumbwbtrecoil} one has:
\begin{eqnarray*}
\lambda&=&2\,\frac{\beta_{0}}{\pi}\,\alpha_{S}L\;,\\
h_{1}\left(\lambda\right)&=&\frac{2\,C_{F}}{\lambda\beta_{0}}\left(\ln\left(1-\lambda\right)+\lambda\right)\;,\\
h_{2}\left(\lambda\right)&=&-\frac{C_{F}\,K}{2\beta_{0}^{2}}\left(\ln\left(1-\lambda\right)+\frac{\lambda}{1-\lambda}\right)-\frac{3\,C_{F}}{2\beta_{0}}\ln\left(1-\lambda\right)\nonumber\\
&&+\frac{C_{F}\beta_{1}}{\beta_{0}^{3}}\left(\frac{1}{2}\ln^{2}\left(1-\lambda\right)+\frac{\ln\left(1-\lambda\right)}{1-\lambda}+\frac{\lambda}{1-\lambda}\right)\;,\\
\Sigma_{s}\left(y\right)&=&e^{L\,h_{1}\left(\lambda\right)+h_{2}\left(\lambda\right)}\;,\\
\mathcal{F}\left(R'\right)&=&\left[\int_{1}^{\infty}\frac{dx}{x^{2}}\left(\frac{1+x}{4}\right)^{-R'}\right]^{2}\frac{e^{-2\gamma_{E}R'}}{\Gamma\left(1+\,R'\right)^{2}}\nonumber\\
&=&\left[\frac{4^{R'}\;{}_{2}F_{1}\left(R',\,1+R';\,2+R';\,-1\right)}{\left(1+R'\right)}\right]^{2}\frac{e^{-2\gamma_{E}R'}}{\Gamma\left(1+\,R'\right)^{2}}\;.
\end{eqnarray*}
These yield:
\begin{eqnarray*}
g_{1}\left(\alpha_{S}L\right)&=&\,h_{1}\left(\frac{\beta_{0}}{\pi}\,\alpha_{S}L\right)\;,\\
g_{2}\left(\alpha_{S}L\right)&=&\,h_{2}\left(\frac{\beta_{0}}{\pi}\,\alpha_{S}L\right)-2\,\ln\left[\Gamma\left(1+R'\right)\right]-2\gamma_{E}R'\nonumber\\
&&+2\,\ln\left[\frac{4^{R'}\;{}_{2}F_{1}\left(R',\,1+R';\,2+R';\,-1\right)}{\left(1+R'\right)}\right]\;.
\end{eqnarray*}
\subsection{Two-to-three Jet Transition in the Durham Algorithm}
From \cite{resumy3a,resumy3b} one has:
\begin{eqnarray*}
\lambda&=&\frac{\beta_{0}}{\pi}\,\alpha_{S}L\;,\\
h_{1}\left(\lambda\right)&=&\frac{C_{F}}{\lambda\beta_{0}}\left(\ln\left(1-\lambda\right)+\lambda\right)\;,\\
h_{2}\left(\lambda\right)&=&-\frac{3C_{F}}{2\beta_{0}}\ln\left(1-\lambda\right)-\frac{C_{F}\,K}{2\beta_{0}^{2}\left(1-\lambda\right)}\left(\lambda+\left(1-\lambda\right)\ln\left(1-\lambda\right)\right)+\nonumber\\
&&+\frac{C_{F}\beta_{1}}{\beta_{0}^{3}}\left(\frac{\lambda+\ln\left(1-\lambda\right)}{1-\lambda}\,+\,\frac{1}{2}\,\ln^{2}\left(1-\lambda\right)\right)\;,\\
\Sigma_{s}\left(y\right)&=&e^{L\,h_{1}\left(\lambda\right)+h_{2}\left(\lambda\right)}\;.
\end{eqnarray*}
The function $\mathcal{F}\left(R'\right)$ for $Y_{3}$ is
known only numerically \cite{resumy3a,resumy3b}. We
interpolate the points using a slightly modified version of
Newton's divided difference formula implemented in the CERN
Computer Program Library. These yield:
\begin{eqnarray*}
g_{1}\left(\alpha_{S}L\right)&=&\,h_{1}\left(\frac{\beta_{0}}{\pi}\,\alpha_{S}L\right)\;,\\
g_{2}\left(\alpha_{S}L\right)&=&\,h_{2}\left(\frac{\beta_{0}}{\pi}\,\alpha_{S}L\right)+\ln\left[\mathcal{F}\left(R'\right)\right]\,.
\end{eqnarray*}

\begin{table}[t]
{\scriptsize
Thrust: $y=\tau=1-T$ and $C$-parameter: $y=C/6$\\
\begin{displaymath}
\begin{array}{ccl}
\hline \\[-1.4ex]
  G_{11} & = & 3\,C_{F} \\[1.2mm]
  G_{12} & = & -2\,C_{F} \\[1.2mm]
  G_{22} & = & \frac{1}{36}\,C_{F}\left(-169C_{A}+22N_{F}+12\left(C_{A}-4C_{F}\right)\pi^{2}\right) \\[1.2mm]
  G_{23} & = & \frac{1}{3}\,C_{F}\left(-11C_{A}+2N_{F}\right) \\[1.2mm]
  G_{33} & = & \frac{1}{108}\,C_{F}\left[-612C_{A}^{2}+180C_{A}N_{F}+108C_{F}N_{F}+\left(11C_{A}-2N_{F}\right)\left(-235\,C_{A}+34\,N_{F}+12\left(C_{A}-6\,C_{F}\right)\pi^{2}\right)+2304\,C_{F}^{2}\zeta(3)\right]\\[1.2mm]
  G_{34} & = & -\frac{7}{108}C_F\left(11\,C_A-2\,N_F\right)^{2} \\[1.2mm]
\hline
\end{array}
\end{displaymath}
Heavy jet mass: $y=\rho$ \\
\begin{displaymath}
\begin{array}{ccl}
\hline\\[-1.4ex]
  G_{11} & = & 3\,C_{F} \\[1.2mm]
  G_{12} & = & -2\,C_{F} \\[1.2mm]
  G_{22} & = & \frac{1}{36}\,C_{F}\left(-169C_{A}+22N_{F}+12\left(C_{A}-2C_{F}\right)\pi^{2}\right) \\[1.2mm]
  G_{23} & = & \frac{1}{3}\,C_{F}\left(-11C_{A}+2N_{F}\right) \\[1.2mm]
  G_{33} & = & \frac{1}{108}\,C_{F}\left[-612C_{A}^{2}+180C_{A}N_{F}+108C_{F}N_{F}+\left(11C_{A}-2N_{F}\right)\left(-235\,C_{A}+34\,N_{F}+12\left(C_{A}-3\,C_{F}\right)\pi^{2}\right)+576\,C_{F}^{2}\zeta(3)\right]\\[1.2mm]
  G_{34} & = & -\frac{7}{108}C_F\left(11\,C_A-2\,N_F\right)^{2} \\[1.2mm]
\hline
\end{array}
\end{displaymath}
Total jet broadening: $y=B_T$ \\
\begin{displaymath}
\begin{array}{ccl}
\hline\\[-1.4ex]
  G_{11} & = & 6C_{F}\\[1.2mm]
  G_{12} & = & -4C_{F}\\[1.2mm]
  G_{22} & = & -\frac{1}{9}C_{F}\left(35C_{A}-2N_{F}-6C_{A}\pi^{2}+24C_{F}\pi^{2}+288C_{F}\,\ln^2 2\right)\\[1.2mm]
  G_{23} & = & -\frac{8}{9}C_{F}\left(11C_{A}-2N_{F}\right)\\[1.2mm]
  G_{33} & = & \frac{2}{81}C_{F}\left(-2471C_{A}^{2}+760C_{A}N_{F}+108C_{F}N_{F}-44N_{F}^{2}+132C_{A}^{2}\pi^{2}-792C_{A}C_{F}\pi^{2}-24C_{A}N_{F}\pi^{2}+144C_{F}N_{F}\pi^{2}\right.\\
&& \qquad\quad\left.+864C_{F}^{2}\pi^{2}\ln 2-9504C_{A}C_{F}\ln^2 2+1728C_{F}N_{F}\ln^2 2-5184C_{F}^{2}\ln^3 2+2376C_{F}^{2}\zeta\left(3\right)\right)\\[1.2mm]
  G_{34} & = & -\frac{2}{9}C_{F}\left(11C_{A}-2N_{F}\right)^{2}\\[1.2mm]
\hline
\end{array}
\end{displaymath}
Wide jet broadening: $y=B_W$ \\
\begin{displaymath}
\begin{array}{ccl}
\hline\\[-1.4ex]
  G_{11} & = & 6C_{F}\\[1.2mm]
  G_{12} & = & -4C_{F}\\[1.2mm]
  G_{22} & = & -\frac{1}{9}C_{F}\left(35C_{A}-2N_{F}-6C_{A}\pi^{2}+288C_{F}\,\ln^2 2\right)\\[1.2mm]
  G_{23} & = & -\frac{8}{9}C_{F}\left(11C_{A}-2N_{F}\right)\\[1.2mm]
  G_{33} & = & \frac{2}{81}C_{F}\left(-2471C_{A}^{2}+760C_{A}N_{F}+108C_{F}N_{F}-44N_{F}^{2}+132C_{A}^{2}\pi^{2}-24C_{A}N_{F}\pi^{2}+864C_{F}^{2}\pi^{2}\ln 2-9504C_{A}C_{F}\ln^2 2\right.\\
&&\qquad\quad\left.+1728C_{F}N_{F}\ln^2 2-5184C_{F}^{2}\ln^3 2-2808C_{F}^{2}\zeta\left(3\right)\right)\\[1.2mm]
  G_{34} & = &  -\frac{2}{9}C_{F}\left(11C_{A}-2N_{F}\right)^{2}\\[1.2mm]
\hline
\end{array}
\end{displaymath}
Two-to-three jet transition in Durham algorithm: $y=Y_3$ \\
\begin{displaymath}
\begin{array}{ccl}
\hline\\[-1.4ex]
  G_{11} & = & 3C_{F}\\[1.2mm]
  G_{12} & = & -C_{F} \\[1.2mm]
  G_{22} & = & \frac{1}{36}C_{F}\left(-35C_{A}+144C_{F}\mathcal{F}_{2}+2N_{F}+6C_{A}\pi^{2}\right) \\[1.2mm]
  G_{23} & = & -\frac{1}{9}C_{F}\left(11C_{A}-2N_{F}\right) \\[1.2mm]
  G_{33} & = &
  \frac{1}{324}C_{F}\left(-2471C_{A}^{2}+4752C_{A}C_{F}\mathcal{F}_{2}+2592C_{F}^{2}\mathcal{F}_{3}+760C_{A}N_{F}+108C_{F}N_{F}-864C_{F}\mathcal{F}_{2}N_{F}-44N_{F}^{2}+132C_{A}^{2}\pi^{2}\right.\\
  &&\left.-24C_{A}N_{F}\pi^{2}\right)\\[1.2mm]
  G_{34} & = & -\frac{1}{72}C_{F}\left(11C_{A}-2N_{F}\right)^{2}\\[1.2mm]
\hline
\end{array}
\end{displaymath}}
\caption{The logarithmic coefficients $G_{ij}$ for LL and
NLL up to the third order in $\alpha_{S}$.}\label{tab:coeff}
\end{table}
\newpage

\begin{table}[t]
\begin{displaymath}
\begin{array}{|c|c|c|c|c|c|c|}
\hline
           & G_{11} & G_{12}   & G_{22}   & G_{23}   & G_{33}   & G_{34} \\
\hline
  \tau /C  & 4.0    & -2.66667 & -24.9388 & -10.2222 & -285.055 & -45.716 \\
\hline
  \rho     & 4.0    & -2.66667 & -13.2415 & -10.2222 & -196.125 & -45.716 \\
\hline
  B_{T}    & 8.0    & -5.33333 & -61.8768 & -27.2593 & -824.787  & -156.741 \\
\hline
  B_{W}    & 8.0    & -5.33333 & -15.0876 & -27.2593 & -472.065 & -156.741 \\
\hline
  Y_{3}    & 4.0    & -1.33333 & 0.867972 & -3.40741 & -28.1784 & -9.7963 \\
\hline
\end{array}
\end{displaymath}
\caption{The numerical value of the logarithmic
coefficients $G_{ij}$ for LL and NLL up to the third order
in $\alpha_{S}$.}\label{tab:numericalcoeff}
\end{table}
\clearpage

\newpage

\newpage

\clearpage


{\small
\begin{thebibliography}{99}

\bibitem{aleph}
  D.~Buskulic {\it et al.}  [ALEPH Collaboration],
  Z.\ Phys.\  C {\bf 73} (1997) 409;\\
  A.~Heister {\it et al.}  [ALEPH Collaboration],
  Eur.\ Phys.\ J.\  C {\bf 35} (2004) 457.

\bibitem{delphi}
  P.~Abreu {\it et al.}  [DELPHI Collaboration],
  Phys.\ Lett.\  B {\bf 456} (1999) 322;\\
  J.~Abdallah {\it et al.}  [DELPHI Collaboration],
  Eur.\ Phys.\ J.\  C {\bf 29} (2003) 285
  [hep-ex/0307048];\\
  J.~Abdallah {\it et al.}  [DELPHI Collaboration],
  Eur.\ Phys.\ J.\  C {\bf 37} (2004) 1
  [hep-ex/0406011].

\bibitem{l3}
  M.~Acciarri {\it et al.}  [L3 Collaboration],
  Phys.\ Lett.\  B {\bf 371} (1996) 137;\\
  M.~Acciarri {\it et al.}  [L3 Collaboration],
  Phys.\ Lett.\  B {\bf 404} (1997) 390;\\
  M.~Acciarri {\it et al.}  [L3 Collaboration],
  Phys.\ Lett.\  B {\bf 444} (1998) 569;\\
  P.~Achard {\it et al.}  [L3 Collaboration],
  Phys.\ Lett.\  B {\bf 536} (2002) 217
  [hep-ex/0206052];\\
  P.~Achard {\it et al.}  [L3 Collaboration],
  Phys.\ Rept.\  {\bf 399} (2004) 71
  [hep-ex/0406049].

\bibitem{opal}
  P.~D.~Acton {\it et al.}  [OPAL Collaboration],
  Z.\ Phys.\  C {\bf 59} (1993) 1;\\
  G.~Alexander {\it et al.}  [OPAL Collaboration],
  Z.\ Phys.\  C {\bf 72} (1996) 191;\\
  K.~Ackerstaff {\it et al.}  [OPAL Collaboration],
  Z.\ Phys.\  C {\bf 75} (1997) 193;\\
  G.~Abbiendi {\it et al.}  [OPAL Collaboration],
  Eur.\ Phys.\ J.\  C {\bf 16} (2000) 185
  [hep-ex/0002012];\\
  G.~Abbiendi {\it et al.}  [OPAL Collaboration],
  Eur.\ Phys.\ J.\  C {\bf 40} (2005) 287
  [hep-ex/0503051].

\bibitem{farhi}
S.~Brandt, C.~Peyrou, R.~Sosnowski and A.~Wroblewski,
  Phys.\ Lett.\  {\bf 12} (1964) 57;\\
E.~Farhi,
  Phys.\ Rev.\ Lett.\  {\bf 39} (1977) 1587.

\bibitem{mh}
L.~Clavelli and D.~Wyler,
  Phys.\ Lett.\  B {\bf 103} (1981) 383.

\bibitem{bwbt}
   P.E.L.~Rakow and B.R.~Webber,
   Nucl.\ Phys.\  B {\bf 191} (1981) 63.

\bibitem{c}
 G.~Parisi,
  Phys.\ Lett.\  B {\bf 74} (1978) 65;\\
  J.F.~Donoghue, F.E.~Low and S.Y.~Pi,
  Phys.\ Rev.\  D {\bf 20} (1979) 2759.



\bibitem{durham}
  S.~Catani, Y.L.~Dokshitzer, M.~Olsson, G.~Turnock and B.R.~Webber,
  Phys.\ Lett.\ B {\bf 269} (1991) 432;\\
 N.~Brown and W.J.~Stirling,
  Phys.\ Lett.\  B {\bf 252} (1990) 657;
  Z.\ Phys.\  C {\bf 53} (1992) 629;\\
W.J.~Stirling {\it et al.}, Proceedings of the Durham Workshop, J.\ Phys.\ \textbf{G17} (1991) 1567;\\
S.~Bethke, Z.~Kunszt, D.E.~Soper and W.J.~Stirling,
  Nucl.\ Phys.\  B {\bf 370} (1992) 310
  [Erratum-ibid.\  B {\bf 523} (1998) 681].


\bibitem{hasko}
 R.W.L.~Jones, M.~Ford, G.P.~Salam, H.~Stenzel and D.~Wicke,
  JHEP {\bf 0312} (2003) 007 [hep-ph/0312016].



\bibitem{resumall}
  S.~Catani, L.~Trentadue, G.~Turnock and B.R.~Webber,
  Nucl.\ Phys.\  B {\bf 407} (1993) 3.


\bibitem{ERT}
R.K.~Ellis, D.A.~Ross and A.E.~Terrano,
Nucl.\ Phys.\ B {\bf 178} (1981) 421.

\bibitem{kunszt}
 Z.~Kunszt,
  Phys.\ Lett.\  B {\bf 99} (1981) 429;\\
J.A.M.~Vermaseren, K.J.F.~Gaemers and S.J.~Oldham,
  Nucl.\ Phys.\  B {\bf 187} (1981) 301;\\
K.~Fabricius, I.~Schmitt, G.~Kramer and G.~Schierholz, Z.~Phys.~C {\bf
  11} (1981) 315.



\bibitem{event}
Z.\ Kunszt and P.\ Nason, in {\it Z Physics at LEP 1}, CERN Yellow Report
89-08, Vol.~1, p.~373;\\
  W.~T.~Giele and E.W.N.~Glover,
  Phys.\ Rev.\  D {\bf 46} (1992) 1980;\\
 S.~Catani and M.~H.~Seymour,
  Phys.\ Lett.\  B {\bf 378} (1996) 287
  [hep-ph/9602277].


\bibitem{ourt}
 A.~Gehrmann-De Ridder, T.~Gehrmann, E.W.N.~Glover and G.~Heinrich,
  Phys.\ Rev.\ Lett.\  {\bf 99} (2007) 132002
  [arXiv:0707.1285].

\bibitem{our3j}
A.~Gehrmann-De Ridder, T.\ Gehrmann, E.W.N.\ Glover and G.\ Heinrich,
  JHEP {\bf 0711} (2007) 058 [arXiv:0710.0346].


\bibitem{ourevent}
 A.~Gehrmann-De Ridder, T.~Gehrmann, E.W.N.~Glover and G.~Heinrich,
 JHEP {\bf 0712} (2007) 094 [arXiv:0711.4711].

\bibitem{ourjets}
 A.~Gehrmann-De Ridder, T.~Gehrmann, E.W.N.~Glover and G.~Heinrich,
  Phys.\ Rev.\ Lett.\ {\bf 100} (2008) 172001
  [arXiv:0802.0813.]



\bibitem{resumt}
S.~Catani, G.~Turnock, B.R.~Webber and L.~Trentadue,
  Phys.\ Lett.\  B {\bf 263} (1991) 491.

\bibitem{resumrho}
 S.~Catani, G.~Turnock and B.R.~Webber,
  Phys.\ Lett.\  B {\bf 272} (1991) 368;\\
  E.~Gardi and J.~Rathsman,
  Nucl.\ Phys.\  B {\bf 638} (2002) 243
  [hep-ph/0201019].

\bibitem{resumbwbt}
 S.~Catani, G.~Turnock and B.R.~Webber,
  Phys.\ Lett.\  B {\bf 295} (1992) 269.

\bibitem{resumbwbtrecoil}
 Y.L.~Dokshitzer, A.~Lucenti, G.~Marchesini and G.P.~Salam,
  JHEP {\bf 9801} (1998) 011
  [hep-ph/9801324].

\bibitem{resumc}
S.~Catani and B.~R.~Webber,
  Phys.\ Lett.\  B {\bf 427} (1998) 377
  [hep-ph/9801350];\\
  E.~Gardi and L.~Magnea,
  JHEP {\bf 0308} (2003) 030
  [hep-ph/0306094].

\bibitem{resumcshoulder}
  S.~Catani and B.R.~Webber,
  JHEP {\bf 9710} (1997) 005
  [hep-ph/9710333].

\bibitem{resumy3a}
  A.~Banfi, G.P.~Salam and G.~Zanderighi,
  JHEP {\bf 0201} (2002) 018
  [hep-ph/0112156].

\bibitem{resumy3b}
  A.~Banfi, G.P.~Salam and G.~Zanderighi,
  JHEP {\bf 0503} (2005) 073
  [hep-ph/0407286].

\bibitem{eecdg}
 D.~de Florian and M.~Grazzini,
  Nucl.\ Phys.\  B {\bf 704} (2005) 387
  [hep-ph/0407241].

\bibitem{scet}
 S.~Fleming, A.~H.~Hoang, S.~Mantry and I.~W.~Stewart,
  Phys.\ Rev.\  D {\bf 77} (2008) 074010
  [arXiv:hep-ph/0703207];\\
S.~Fleming, A.~H.~Hoang, S.~Mantry and I.~W.~Stewart,
  arXiv:0711.2079;\\
M.D.~Schwartz,
  Phys.\ Rev.\  D {\bf 77} (2008) 014026
  [arXiv:0709.2709];\\
C.W.~Bauer, S.P.~Fleming, C.~Lee and G.~Sterman,
  arXiv:0801.4569.


\bibitem{scetthrust}
T.~Becher and M.D.~Schwartz, arXiv:0803.0342.


\bibitem{giulia}
A.\ Banfi and G.\ Zanderighi, private communication.

\bibitem{bethke}
 S.~Bethke,
  Prog.\ Part.\ Nucl.\ Phys.\  {\bf 58} (2007) 351
  [hep-ex/0606035].

\bibitem{ouralphas}
 G.\ Dissertori,
A.~Gehrmann-De Ridder, T.~Gehrmann, E.W.N.~Glover, G.~Heinrich and
H.\ Stenzel, JHEP {\bf 0802} (2008) 040
 [arXiv:0712.0327].



\bibitem{quarkmass}
 W.~Bernreuther, A.~Brandenburg and P.~Uwer,
  Phys.\ Rev.\ Lett.\  {\bf 79} (1997) 189
  [hep-ph/9703305];\\
 A.~Brandenburg and P.~Uwer,
  Nucl.\ Phys.\  B {\bf 515} (1998) 279
  [hep-ph/9708350];\\
 G.~Rodrigo, A.~Santamaria and M.~S.~Bilenky,
  Phys.\ Rev.\ Lett.\  {\bf 79} (1997) 193
  [hep-ph/9703358];\\
  P.~Nason and C.~Oleari,
  Nucl.\ Phys.\  B {\bf 521} (1998) 237
  [hep-ph/9709360].






\end{thebibliography} }
\end{document}